\documentclass[prl,aps,groupedaddress,floatfix,twocolumn,superscriptaddress,showpacs]{revtex4-1}
\usepackage{amsmath,amssymb,multirow,epsfig,bm,pifont}
\usepackage[linkcolor=black,citecolor=black,urlcolor=blue,colorlinks=true,linktocpage=true]{hyperref}
\usepackage{graphicx}
\usepackage{epsfig}
\usepackage{bm}
\usepackage{tabularx}
\usepackage{makecell}

\newcommand{\beq}{\begin{equation}}
\newcommand{\eeq}{\end{equation}}
\newcommand{\bea}{\begin{eqnarray}}
\newcommand{\eea}{\end{eqnarray}}
\newcommand{\tr}{\mathrm{tr}}

\begin{document}

\title{Convexity of the entanglement energy of SU($2N$)-symmetric \\
fermions with attractive interactions}

\author{Joaqu\'{\i}n E. Drut}
\email{drut@email.unc.edu}
\affiliation{Department of Physics and Astronomy, University of North Carolina, Chapel Hill, NC, 27599, USA}

\author{William J. Porter}
\email{wjporter@live.unc.edu}
\affiliation{Department of Physics and Astronomy, University of North Carolina, Chapel Hill, NC, 27599, USA}

\begin{abstract}
The positivity of the probability measure of attractively interacting systems of $2N$-component fermions
enables the derivation of an exact convexity property for the ground-state energy of such systems. 
Using analogous arguments, applied to path-integral expressions for the $n$-th R\' enyi entanglement entropy $S^{}_n$ derived recently, 
we prove non-perturbative analytic relations for the entanglement energies $\mathcal E^{}_n$ of those systems defined via 
$\mathcal{E}^{}_n \equiv \frac{n-1}{n}T S^{}_n + F$ where $\beta = 1/T$ is the extent of the imaginary time direction and 
$-\beta F = \ln \mathcal Z$ where $\mathcal Z$ is the partition sum appropriate to the temperature.
These relations are valid for all sub-system sizes, particle numbers and dimensions, and in arbitrary external
trapping potentials.
\end{abstract}

\date{\today}
\maketitle

%%%%%%%%%%%%%%%%%%%%%%%%%%%%%%%%%%%%%%%%%%%%%
\emph {Introduction.--} 
In recent years, research in many-body and condensed matter physics in general has developed 
a growing intersection with the areas of quantum information and quantum computation~\cite{RevModPhys}. 
Considerable attention has been given to the notion of entanglement, in particular to the entanglement entropies
as a means to identify and characterize quantum phase transitions (especially topological ones)~\cite{Kitaev,Hertzberg:2012mn}.
The conformal field theories governing the low-energy behavior at those critical points are also 
of relevance for high-energy physics, black-hole physics and string theory~\cite{HEP}, where some exact
results have been available for many years.

In spite of much progress, the calculation of the entanglement properties
has been a difficult challenge for many-body theories. Indeed, even the calculation of the asymptotic form of the 
entanglement entropy for large (sub-)systems of non-interacting fermions was the subject of much discussion
(see e.g.~\cite{LargeLNonInt}); and for interacting systems it
was until recently challenging to even formulate a way amenable to quantum Monte Carlo calculations.
However, there are currently a few alternatives (see e.g.~\cite{QMC}), and in this work we will use the formalism of Ref.~\cite{Grover}, in which 
Grover proposes a way to calculate fully interacting reduced density matrices using auxiliary-field quantum Monte Carlo (AFQMC)
(see also Ref.~\cite{Assaad}). The resulting path-integral expression for the $n$-th R\' enyi entanglement entropy will be central to our arguments.

Such AFQMC calculations, in particular those on the lattice, rely on the positivity of the fermion determinant 
as the natural probability measure arising from the field integration of the fermion degrees of freedom.
While this property is essential from a computational standpoint, it has also enabled the non-perturbative analytic proof of 
a variety of exact inequalities among the hadron masses arising from Quantum Chromodynamics (QCD); the most recent 
contribution in this direction (as of this writing) is due to Detmold~\cite{DetmoldInequalityQCD}, but earlier
versions have been around for many years~\cite{GeneralQCDInequalities} (see Ref.~\cite{GeneralQCDInequalitiesReview}
for a review on QCD inequalities).

\begin{figure}[h]
\includegraphics[width=0.8\columnwidth]{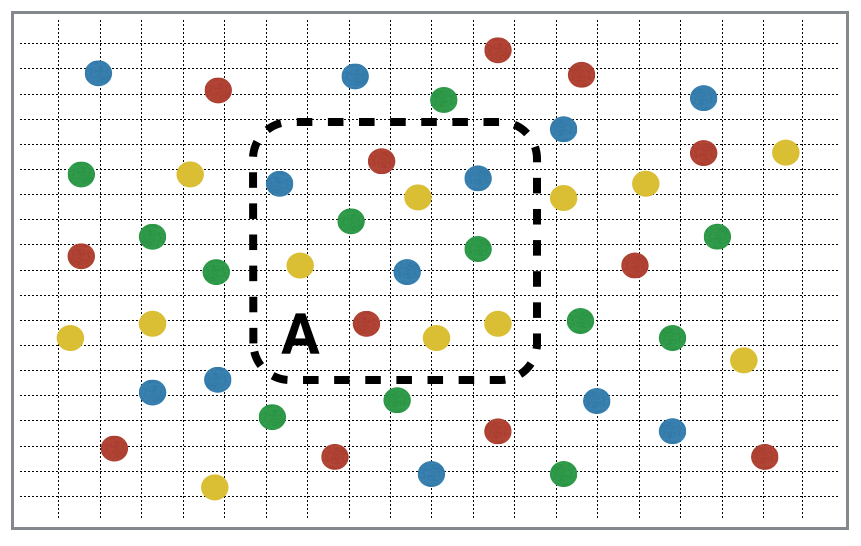}
\caption{\label{Fig:MFS}(color online) The system considered here is made out of $2N$ species of otherwise
identical fermions, on a space-time lattice. All of the species live in the whole volume, but the entanglement entropy
is computed for a subregion $A$.}
\end{figure}

In this work, we use measure-positivity arguments to derive an exact inequality that expresses the convexity 
of the entanglement energy $\mathcal{E}^{}_n$ (defined below) with respect to particle number in a system of SU($2N$) fermions, i.e. a system of $2N$ 
species of otherwise identical fermions, with attractive interactions (depicted in Fig.~\ref{Fig:MFS}).  For a positive integer $n$, we define the entanglement energy as
\beq
\mathcal{E}^{}_n \equiv \frac{n-1}{n}T S^{}_n + F
\eeq
where we denote by $\beta = 1/T$ the extent of the imaginary time direction and by $\mathcal Z$ the zero- or finite-temperature partition sum with $-\beta F = \ln \mathcal Z$.
The inequality we obtain holds for the energy at all orders, all sub-system sizes, 
particle numbers and dimensions, as well as arbitrary external trapping potentials.

\begin{figure}[h]
\includegraphics[width=1.0\columnwidth]{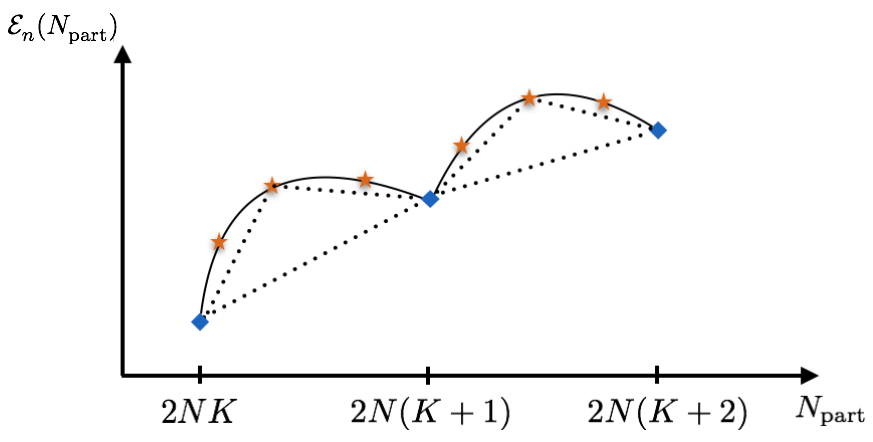}
\caption{\label{Fig:SnConvex}(color online) Convexity of the entanglement energy $\mathcal{E}^{}_n$ 
as a function of particle number, for a system of $2N$ species. The convexity property is valid 
for systems with particle numbers between $K$ and $K+1$ particles per component, with $K\!\in\!\mathbb{N}$, 
such that some components have $K$ particles and the rest $K+1$. The property holds for arbitrary region 
$A$, R\'enyi order $n$, spatial dimension, and external potential.}
\end{figure}

To carry out the proof, we combine a path-integral expression for the R\'enyi entanglement
entropy of interacting Fermi systems, derived recently in Ref.~\cite{Grover}, with (the proof of) a theorem 
that establishes the spectral convexity of SU($2N$)-symmetric fermions with attractive interactions, 
first proven by D. Lee in Ref.~\cite{Lee}. Interestingly, this theorem was originally motivated by the properties of 
nuclear structure~\cite{LeeEtAl} and the importance of Wigner SU(4) symmetry in that context~\cite{Wigner}.
To facilitate contact with the original derivation, we 
follow closely the main steps, and use similar notation. However, some modification is needed
due to the presence of ``replicas'' of the auxiliary field, as explained below, among other details.
The main result is
depicted in Fig.~\ref{Fig:SnConvex}, and it indicates that the entanglement entropy of $N_\text{part}$
fermions, for $2N K \leq N_\text{part} \leq 2N (K+1)$, with $K$ a positive integer, is always greater or equal than 
the average of the entanglement entropies of $2N K$ and $2N (K+1)$ fermions. Here, the $2N K$ fermions are to be understood
as $K$ particles per component, and the $2N (K+1)$ fermions appear as $K + 1$ particles per component.
The states between these have $K + 1$ particles for some species and $K$ particles for the rest.
It is well-known that other inequalities exist that 
relate the entanglement entropy of disjoint subregions of a system (see e.g.~\cite{OtherInequality}). 
The inequality established here, however, is unrelated to the property of sub-additivity, 
and is therefore of a different kind, as explained below.

%%%%%%%%%%%%%%%%%%%%%%%%%%%%%%%%%%%%%%%%%%%%%%
\emph {Derivation of the inequality.--}
The $n$-th R\'enyi entanglement entropy $S^{}_{n}$ of a sub-system $A$ of a given quantum
system is defined by
\beq
S^{}_{n} = \frac{1}{1-n} \ln \tr (\hat \rho_A^n),
\eeq
where $\hat  \rho^{}_A$ is the reduced density matrix of sub-system $A$ 
(i.e. the degrees of freedom of the rest of the system are traced over).

To establish our main result, we put the system on a $d$-dimensional spatial lattice of side $N_x^{}$
and consider implementing the projection AFQMC method, with some trial state $| \Psi \rangle$, which we assume has a 
non-vanishing overlap with the true ground state. 

In Ref.~\cite{Grover}, Grover derived an auxiliary-field path-integral form for $\hat \rho_A$, from which 
he showed that $S^{}_{n}$ can be accessed via AFQMC calculations. 
Indeed, for a system of $2N$-component fermions, $S^{}_{n}$ can be written in that formalism as
\bea
\label{Eq:Sn}
\exp\text{\big(}(1-n)S^{}_{n} \text{\big)} =
\int \mathcal D {\{\sigma\}}^{} P[\{\sigma \}]\; Q[\{\sigma \}]
\eea
where 
\beq
\mathcal D {\{\sigma\}}^{}  = \prod_{k=1}^{n} \frac{\mathcal D {\sigma^{}_k}}{\mathcal Z},
\eeq
\beq
\mathcal Z = \int \mathcal D {\sigma}^{}\; \prod_{m=1}^{2N} {{\det}U^{}_{m}[\sigma]},
\eeq
\beq
P[\{\sigma \}] = \prod_{k=1}^{n} \prod_{m=1}^{2N} {{\det}U^{}_{m}[\sigma^{}_k]},
\eeq
\beq
Q[\{\sigma \}] = \prod_{m=1}^{2N} {{\det}M^{}_{m}[\{\sigma \}]}
\eeq
\bea
M_{m}[\{\sigma\}] &=& \prod_{k=1}^{n} \left(\openone - G^{}_{A,m}[\sigma^{}_k]\right)\times \nonumber \\ 
&&\left[\openone + \prod_{k=1}^{n}\frac{G^{}_{A,m}[\sigma^{}_k]}{\openone - G^{}_{A,m}[\sigma^{}_k]} \right],
\eea
and the path integral $\int \mathcal D {\{\sigma\}}^{}$ is over the $n$ ``replicas'' $\sigma^{}_k$ of the 
Hubbard-Stratonovich auxiliary field.
Here, $U_m[\sigma]$ is a matrix which encodes the dynamics of the $m$-th component in 
the system, namely the kinetic energy and the form of the interaction after a Hubbard-Stratonovich transformation;
it also encodes the form of the trial state $|\Psi \rangle$ (see e.g. Ref.~\cite{reviewMC}), which we take to be a Slater
determinant. $G^{}_{A,m}[\sigma^{}_k]$ is the restricted non-interacting single-particle Green's 
function of the $m$-th component in the background field $\sigma^{}_k$, as explained in Ref.~\cite{Grover} 
(see also Ref.~\cite{Peschel}). The size of $U_m[\sigma]$ is given by the number of particles of
the $m$-th species present in the system. The size of $G^{}_{A,m}[\sigma^{}_k]$, on the other hand, is
given by the number of lattice sites enclosed by the region $A$.
Note that, separating a factor of $\mathcal Z^n$ in the denominator of Eq.~\ref{Eq:Sn}, an explicit form 
can be identified in the numerator as the result of the so-called ``replica trick" (namely a partition function for 
$n$ copies of the system, ``glued'' together in the region $A$).
%As is well known, when an even number $2N$ of flavors is considered, and the interactions are attractive, 
%then $\det^{2N} U[\sigma]$ is real and positive for all $\sigma$, which means that $P[\{\sigma \}]$ is
%a well-defined probability measure.

\begin{figure}[h]
\includegraphics[width=1.0\columnwidth]{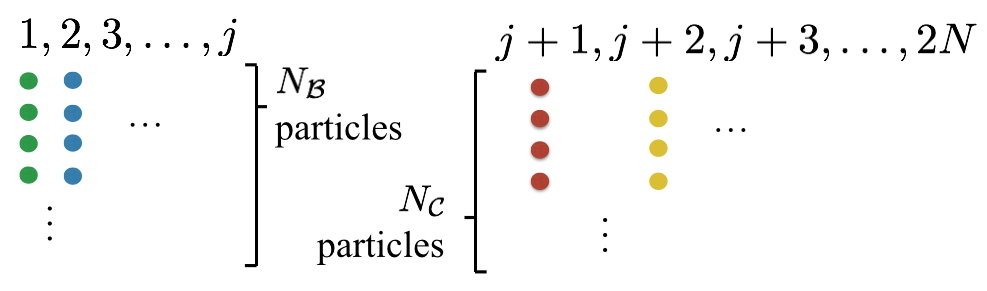}
\caption{\label{Fig:NbNc}(color online) Distribution of particles among $2N$ fermion species for the proof of 
the inequality.}
\end{figure}

Because we are considering a finite lattice, the single-particle space is of finite size $N_x^d$.
As in Ref.~\cite{Lee}, we may define our trial state $|\Psi\rangle$ as a Slater determinant, letting $j$ components fill 
some $N_{\mathcal B}$ single-particle orbitals in a subset $\mathcal B$ (the same set of orbitals for all those $j$ 
components), and letting the remaining $2N-j$ components fill some other set $\mathcal C$ of, say, $N_{\mathcal C}$ 
orbitals (see Fig.~\ref{Fig:NbNc}). The total number of fermions is therefore $N_\text{part} = j N_{\mathcal B} + (2N-j) N_{\mathcal C}$. 

With this trial state, the matrix $U_m$ will be the same for species $1,\dots,j$ and we shall denote it $U_{\mathcal B}$;
similarly, $U_m$ will be the same for species $j+1,\dots,2N$ and we shall denote it $U_{\mathcal C}$.
Therefore,
\bea
P[\{\sigma \}] &\!=\!& 
%\prod_{k=1}^{n} {(\det U_{\mathcal B}[\sigma_k]})^{j}
%({\det U_{\mathcal C}[\sigma_k]})^{2N - j} \nonumber \\
%&\!=\!&  
\prod_{k=1}^{n} {(\det U_{\mathcal B}[\sigma_k]})^{j} 
\prod_{k=1}^{n} {(\det U_{\mathcal C}[\sigma_k]})^{2N - j},
\eea
and 
\bea
Q[\{\sigma \}] &\!=\!& 
{(\det M_{\mathcal B}[\{\sigma\}]})^{j}
({\det M_{\mathcal C}[\{\sigma\}]})^{2N - j}.
\eea

In order to prove the advertised convexity property, we use H\"older's inequality, which
states that
\beq
\label{Eq:Hoelder}
\int dx | f(x) g(x) | \leq \left [\int dx |f(x)|^{p} \right ]^{1/p} \left [ \int dx  | g(x) |^q \right ]^{1/q},
\eeq
for any functions $f(x)$, $g(x)$ for which the integral exists, and where $1/p + 1/q = 1$.
We apply this identity to our path integral expression for $\exp\text{\big(}(1-n)S^{}_{n}\text{\big)}$ in Eq.~\ref{Eq:Sn}. 
Before doing that, the following should be noted. This separation of components into $j$ and $2N-j$ implies 
that the integrand on the right-hand-side of Eq.~\ref{Eq:Sn} is not in general positive definite (unless $j$ is even
and only if the determinants are real). 
However, the integral of Eq.~\ref{Eq:Sn} is naturally bounded from above by the integral over the {\it absolute value} of the integrand,
which is what we will identify as the left-hand side of Eq.~\ref{Eq:Hoelder}.

For this purpose, we choose integers $n^{}_1, n^{}_2$ such that $0 \leq 2 n^{}_1 \leq j \leq 2 n^{}_2 \leq 2N$, and take the integration measure 
$dx$ to be $\mathcal D \{\sigma\} {\tilde P}[\{\sigma \}]$, where
\bea
\!\!\!\!
{\tilde P}[\{\sigma \}]\!=\!
 {(\det M^{}_{\mathcal B}[\{\sigma\}]})^{2 n^{}_1} ({\det M^{}_{\mathcal C}[\{\sigma\}]})^{2N - 2 n^{}_2} \times 
 \nonumber \\
\prod_{k=1}^{n} {(\det U^{}_{\mathcal B} [\sigma^{}_k]})^{2 n^{}_1} ({\det U^{}_{\mathcal C}[\sigma^{}_k]})^{2N - 2n^{}_2} ,
\eea
which is positive definite as long as the determinants are real. This can be guaranteed for attractive interactions.
We take, furthermore,
\bea
\!\!\!\!\!\!
|f(x)| &\to& |\det M^{}_{\mathcal B}[\{\sigma\}]|^{j-2n^{}_1} \prod_{k=1}^{n} { |\det U^{}_{\mathcal B}[\sigma^{}_k]}|^{j-2n^{}_1}, \\
\!\!\!\!\!\!
|g(x)| &\to& |\det M^{}_{\mathcal C}[\{\sigma\}]|^{2n^{}_2-j} \prod_{k=1}^{n} {| \det U^{}_{\mathcal C}[\sigma^{}_k]}|^{2n^{}_2-j},
\eea
and  
\bea
p = (2n^{}_2 - 2n^{}_1)/({j-2n^{}_1}), \\
q = (2n^{}_2 - 2n^{}_1)/({2n^{}_2 - j}).
\eea

Putting everything together, it is a matter of simple algebra to see that
\bea
\label{Eq:SnInequality}
\mathcal{E}^{}_{n}[{\mathcal B}^{j},{\mathcal C}^{2N - j}] &\geq&
\frac{j-2n^{}_1}{2n^{}_2 - 2n^{}_1} \mathcal{E}^{}_{n}[{\mathcal B}^{2n^{}_2},{\mathcal C}^{2N - 2n^{}_2}]  + \nonumber \\
&&\ \ \ \frac{2n^{}_2 - j}{2n^{}_2 - 2n^{}_1} \mathcal{E}^{}_{n}[{\mathcal B}^{2n^{}_1},{\mathcal C}^{2N - 2n^{}_1}],
\eea
for all $n > 1$, where $\mathcal{E}^{}_{n}[{\mathcal B}^{j},{\mathcal C}^{2N - j}]$ is the estimate of the entanglement
energy obtained with the trial state described above. If the trial state has a non-vanishing overlap
with the ground state of the system, as commonly assumed in AFQMC calculations, then 
$\mathcal{E}^{}_{n}[{\mathcal B}^{j},{\mathcal C}^{2N - j}]$ will converge to the true entanglement energy 
of the system $\mathcal{E}^{}_{n}$ in the limit of large imaginary times.

This is, in essence, the final result. It indicates that $\mathcal{E}^{}_{n}$
is a convex function, in the sense that its values for any $j$ between $2 n_1^{}$ and $2 n_2^{}$ are
larger than the average of the values at those two points.
We note also that the inequality is trivially saturated in the absence of interactions, 
as no path integral is present in Eq.~\ref{Eq:Hoelder} in that case.

%%%%%%%%%%%%%%%%%%%%%%%%%%%%%%%%%%%%%%%%%%%%%%
\emph {Discussion.--} Taking $N_{\mathcal B} = K$ and $N_{\mathcal C} = K + 1$, we see that the total number of fermions is
$N_\text{part} = j N_{\mathcal B} + (2N-j) N_{\mathcal C} = 2NK + j$. Since $0 \leq j \leq 2N$, we see that $\mathcal{E}^{}_{n}$
must be convex between $2NK$ and $2N(K+1)$, for any $K$.
The inequality holds in particular for $K = 0$, which shows that $\mathcal{E}^{}_{n}$ is convex for all particle numbers $0 \leq N_\text{part} \leq 2N$ 
as well. These statements are depicted in Fig.~\ref{Fig:SnConvex}. We stress that these results are valid for arbitrary sub-region 
$A$, R\'enyi order $n$, and spatial dimension. Furthermore, the system may be in an arbitrary external potential, in particular
a harmonic trap such as those routinely used in ultracold atom experiments.

Remarkably, our main result is also valid even if the region $A$ is not a region of coordinate space.
Indeed, the formalism of the proof presented here, itself based on the second-quantization derivations of
Ref.~\cite{Grover}, makes no explicit reference to the space in which the single-particle orbitals are defined.
As a consequence, the inequality Eq.~\ref{Eq:SnInequality} is valid also when $\mathcal{E}^{}_{n}$ is computed in momentum space,
harmonic oscillator space, etc., as long as one can guarantee that ${\det} M_m[\{\sigma\}]$ is real. 

The above proof can be generalized to finite temperature in the grand-canonical ensemble. The role of the matrix $U_m[\sigma_k]$
is now played by its full-sized ($N_x^d \times N_x^d$) finite-temperature form $\mathcal U_m[\sigma_k]$, such that
$U_m[\sigma_k] \to \openone + z_m\mathcal U_m[\sigma_k]$, where $z_m$ is the fugacity of the $m$-th species.
The matrix $M$, on the other hand, retains its definition in terms of reduced Green's functions, but the latter are 
replaced by their finite-temperature counterparts.
The proof then proceeds in the same fashion, but one assigns different chemical potentials $\mu^{}_{\mathcal B}$ to 
components $1\dots j$, and $\mu^{}_{\mathcal C}$ to components $j+1,\dots,2N$, rather than specific particle numbers. 
The total average particle number is then 
$\langle N_\text{part} \rangle = j \langle N_\mathcal{B} \rangle + (2N - j)\langle N_\mathcal{C} \rangle$.
Tuning the chemical potentials such that $\langle N_{\mathcal B} \rangle = K$ and $\langle N_{\mathcal C} \rangle = K + 1$,
we see that $\mathcal{E}^{}_{n}$ must be convex between $2NK$ and $2N(K+1)$, for any $K$, also at finite temperature.

Throughout the proof we implicitly assumed to be considering non-relativistic systems, for which particle number is a 
good quantum number. The generalization of our result to relativistic systems, at least in the absence of gauge fields,
can be expected to be straightforward, with states of fixed particle number replaced by states of fixed charge. Those
systems will be considered elsewhere.

%Although there is no simple expression for the von Neumann entropy in terms of path integrals (that we are aware of), 
%we may still try to make progress in that direction. The von Neumann entropy is 
%%
%\beq
%S^{}_{n} = -\tr (\rho^{}_A \log\rho^{}_A),
%\eeq
%
%We may expand,
%%
%\beq
% \log\rho^{}_A = \sum_{k=1}^{\infty} \frac{(-1)^{k+1}}{k} {(\rho^{}_A-\openone)^k} =
% \sum_{k=1}^{\infty} \frac{(-1)^{k+1}}{k} \sum_q C_{q,k}{\rho^{k}_A}...
%\eeq

%%%%%%%%%%%%%%%%%%%%%%%%%%%%%%%%%%%%%%%%%%%%%%
\emph {Summary and Conclusions.--} In this work, we have connected a recently derived path-integral form of the 
R\'enyi entanglement energy $\mathcal{E}^{}_{n}$ with a theorem originally proven in the context of nuclear structure to show 
that $\mathcal{E}^{}_{n}$ is a convex function when the particle number is varied. The inequality underlying the property of convexity 
is remarkably general, holding for all R\'enyi energies, all sub-system sizes, 
particle numbers and dimensions, as well as arbitrary external trapping potentials. If an analytic continuation in $n$ is
possible then we expect convexity to be maintained for the von Neumann entropy.
The fundamental constraint of the proof is the positivity of the probability
measure, which in turn only depends on the SU($2N$) symmetry and the interaction being attractive.
The inequality holds in particular for the attractive Hubbard model in arbitrary dimensions.
These results can be further generalized to finite temperature and to relativistic systems.

While the experimental determination of the entanglement entropy is currently beyond reach, attempts are 
underway with ultracold atoms, in particular in optical lattices (see e.g.~\cite{experiments}). As is well known, these systems
are remarkably malleable. Most relevant for this work is the study of few-atom systems of atomic species with a varying number of 
internal degrees of freedom, such as Yb and Sr isotopes~\cite{Rey}; these make it possible to
realize SU(6) and SU(10) symmetries, thus going well beyond the usual SU(2) case commonly realized with Li and K atoms.
It is for these systems that the results of this work represent a prediction. 

On the other hand, as pointed out in Ref.~\cite{Lee}, it appears that real nuclei respect the inequalities that would
strictly speaking only be valid in the attractive SU(4)-symmetric limit of degenerate nucleons. It would therefore
be of interest to determine whether the entanglement energy of real nuclei also respects the convexity property.
With recent advances in coupled-cluster and Green's function and lattice Monte Carlo calculations, this may be 
expected to be determined theoretically sooner than experimentally.

%%%%%%%%%%%%%%%%%%%%%%%%%%%%%%% 
\emph {Acknowledgements.--} 
We gratefully acknowledge discussions with D. Lee.
This work was supported in part by the U.S. National Science Foundation under Grant No. PHY1306520.

%%%%%%%%%%%%%%%%%%%%%%%%%%%%%%%
% Bibliography

%%%%%%%%%%%%%%%%%%%%%%%%%%%%%%%%%%%%%%%%%%%%%

\end{document}